\documentclass[a4paper]{aa}
\usepackage{graphicx,float}
\usepackage{latexsym,amsmath,amssymb}
\usepackage{natbib}
\usepackage[scriptsize,tight]{subfigure}
\bibpunct{(}{)}{;}{a}{}{,}

\begin{document}

\title{Probing clumpy stellar winds with a neutron star}

\author{R. Walter\inst{1, 2} \and J. Zurita Heras\inst{3}}
\institute{
INTEGRAL Science Data Centre, Chemin d'Ecogia 16, CH-1290 Versoix, Switzerland, \email{Roland.Walter@obs.unige.ch}
\and
Observatoire de Gen\`eve, Universit\'e de Gen\`eve, Chemin des Maillettes 51, CH-1290 Sauverny, Switzerland
\and
Laboratoire AIM, CEA/DSM-CNRS-Universit\'e Paris Diderot, DAPNIA/Service d'Astrophysique, FR-91191 Gif-sur-Yvette, France, \email{juan-antonio.zurita-heras@cea.fr}
}
\offprints{Roland Walter}

\date{Received 25 July 2007 / Accepted 8 October 2007}

\abstract
{INTEGRAL, the European Space Agency's $\gamma$-ray observatory, tripled the number of super-giant high-mass X-ray binaries (sgHMXB) known in the Galaxy by revealing absorbed and fast transient  (SFXT) systems.}
{In these sources, quantitative constraints on the wind clumping of the massive stars could be obtained from the study of the hard X-ray variability of the compact accreting object.}
{Hard X-ray flares and quiescent emission of SFXT systems have been characterized and used to derive wind clump parameters.}
{A large fraction of the hard X-ray emission is emitted in the form of flares with a typical duration of 3 ks, frequency of 7 days and luminosity of $10^{36}$ erg/s. Such flares are most probably emitted by the interaction of a compact object orbiting at $\sim10~R_*$ with wind clumps ($10^{22-23}$ g) representing a large fraction of the stellar mass-loss rate. The density ratio between the clumps and the inter-clump medium is  $10^{2-4}$ in SFXT systems. 
}
{The parameters of the clumps and of the inter-clump medium, derived from the SFXT flaring behavior, are in good agreement with macro-clumping scenario and line driven instability simulations. SFXT have probably a larger orbital radius than classical sgHMXB.}


\maketitle

\section{Introduction}
\label{sec:intro}

Stellar winds have profound implications for the evolution of massive stars, on the chemical evolution of the Universe and as source of energy and momentum in the interstellar medium. 

Photons emitted by  massive stars directly transfer momentum to the stellar wind through absorption in numerous Doppler shifted optically thick spectral lines \citep{cak1975} and drive the wind to highly supersonic speeds.
Such line driven stellar winds are very unstable \citep{ocr1988,feldmeier1995}. The collisions between high speed and low density material with slower gas trigger strong shocks, form high density contrasts and wind clumps and lead to thermal X-ray emission.

There are multiple observational lines of evidence for wind clumping in massive stars from wind line profiles \citep{Bouret2005,Fullerton2006}, discrete absorption components \citep{PrinjaHowarth1986}, variable line profiles \citep{Lepine1999,Markova2005}, polarimetry \citep{Lupie1987, Davies2007}, 
X-rays continuum \citep{CassinelliOlson1979} and line emission \citep{OskinovaFeldmeierHamann2006}. 

Wind clumping has an important effect on mass-loss rate diagnostics that depends on the square of the wind density. It leads to a reduction of the stellar mass-loss rate estimates by a factor $f_V^{-0.5}$ where $f_V$ is the clump volume filling factor \citep{Abbott1981,Fullerton2006}. A reduction by a factor $>3$ becomes problematic for massive star evolution \citep{Hirschi2007}. Optically thick clumps may however help to reconcile wind clumping and usual mass-loss rates by reducing spectral features and free-free emission \citep{OskinovaHamannFeldmeier2007}.

Indirect measures of the structure of massive star winds are possible in binary systems through the analysis of the interaction between the stellar wind and the companion or its stellar wind. 
In colliding wind binaries, \cite{Schild2004} and \cite{Pollock2005} have pointed out that the observed X-ray column densities are much lower than expected from smooth winds and that clumped winds are the probable solution \citep[see also][]{pittard2007}.
In wind-fed accreting High-Mass X-Ray Binaries (HMXB), 
clumping has been invoked to explain the orbital variability of X-ray line profiles emitted by the photo-ionized wind \citep{Sako2003,Vandermeer2005}. 

In this  paper we aim at studying the clumping of stellar wind in HMXB using the hard X-ray variability observed by the IBIS/ISGRI instrument \citep{ubertini03AA,lebrun03AA} on board the INTErnational Gamma-Ray Astrophysics Laboratory \citep[INTEGRAL,][]{winkler03AA} which could be used to probe clump parameters and density contrasts in the stellar wind. This study follows a first paper  \citep{Leyder2007} interpreting the flaring behavior of \object{IGR J08408$-$4503} in term of wind clumping.

Classical wind-fed super-giant HMXB (sgHMXB) are made of a compact object orbiting within few stellar radii from a super-giant companion. Recently INTEGRAL almost tripled the number of sgHMXB systems known in the Galaxy and revealed a much more complex picture with two additional families of sources: the highly absorbed persistent systems \citep{walter04esasp,walter2006} and the super-giant fast X-ray transient (SFXT) systems \citep{Negueruela2006}. 

The highly absorbed systems have orbital and spin periods similar to those of classical Roche-lobe underflow sgHMXB, however the absorbing column densities are much higher than observed on average in classical systems \citep{walter2006}. The fast transient systems are characterized by fast outbursts, by a low quiescent luminosity and by super-giant OB companions \citep{Sguera2006,Negueruela2007}.

The sample of sources and the INTEGRAL data analysis are described in Sect. 2. Their hard X-ray variability is discussed in the context of clumpy stellar winds in Sect. 3. Finally Sect. 4 summarizes our principal conclusions.

\section{Sources and data analysis}

\subsection{Source sample}

Many sources have now been proposed as candidate super-giant fast X-ray transient based on their hard X-ray variability characteristics, and, for a subset of them, optical counterpart spectral type. Contrasting statements have however been made on specific sources for what concerns their persistent or transient nature. In the frame of the current paper we have considered all SFXT candidates together with several persistent and absorbed super-giant HMXB for comparison. 

Among them, we specifically excluded known Be systems (IGR J11305-6246, AX J1700.2-5129), sources detected only once by INTEGRAL (IGR J16283-4843, IGR J16358-4726),
blended INTEGRAL sources (IGR J17407-2808),
the long period systems IGR J11215-5952 \citep{Sidoli2006} and AX J1749.1-2733 \citep{ZuritaChaty2007} and the sgB[e] system \object{IGR J16318$-$4848} \citep{walter2006}.
 
We analyzed the available INTEGRAL data for 12 candidate SFXT that have large variability factors (table \ref{tab2}) and compared them with classical and absorbed sgHMXB system that have typical variability factor $\lesssim20$.

The 12 classical sgHMXB systems (\object{4U 1700-37}, \object{4U 1538-52}, \object{SAX J1802.7-2017}, \object{Cyg X-1}, \object{XTE J1855-026}, \object{Vela X-1}, \object{2S 014+65}, \object{1E 1145.1-6141}, \object{GX 301-2}, \object{4U0900-40}, \object{4U1907+09}, \object{XTE J1855-026}) have orbital radii ranging between 1.5 and 2.5 stellar radii.

The 7 absorbed sgHMXB systems (\object{IGR J16320$-$4751} \citep{Rodriguez2006}, \object{IGR J16393$-$4641} \citep{bodaghee05}, \object{IGR J17252$-$3616} \citep{zurita05}, \object{IGR J18027$-$2016} \citep{hill05aa}, \object{IGR J19140+0951} \citep{rodriguez05aa}, \object{IGR J16418$-$4532} \citep{walter2006}, \object{IGR J00370+6122} \citep{Zand2007}) have orbital periods $\lesssim15~\rm{days}$ and probable super-giant companions i.e. small orbital radii as well.

\begin{table*}[t]
\caption{List of SFXT candidates with quiescent flux $F_{q}$, flaring characteristics: 
range of flare count rates $F_{fl}$, number of short ($<15$ ks) and long ($>15$ ks) flares $N_{fl}$, the average and range of durations for short flares and the list of durations for long flares $t_{fl}$ and source observing elapsed time $T_{obs}$.}
\label{tab2}
\begin{tabular}[c]{l|c|c|r@{.}l|l|c|l|l|r}
\hline \hline \noalign{\smallskip}
Source                   &Spectral&Distance&\multicolumn{2}{c|}{$F_{q}$}&\multicolumn{1}{c|}{$F_{fl}$}&$N_{fl}$&\multicolumn{1}{c|}{$t_{fl}$ [short]}&\multicolumn{1}{c|}{$t_{fl}$ [long]}&\multicolumn{1}{c}{$T_{obs}$}\\
                  &Type&kpc&\multicolumn{2}{c|}{ct/s}&\multicolumn{1}{c|}{ct/s}&short+long&\multicolumn{1}{c|}{ks}&\multicolumn{1}{c|}{ks}&\multicolumn{1}{c}{days}\\
\noalign{\smallskip\hrule\smallskip}
\noalign{\vspace{2mm} \bf SFXT systems\vspace{2mm}}
\object{IGR\,J08408$-$4503}     &O8.5 Ib$^1$     &2.8$^1$         &$<0$&1&2.1-3.9  &2+0    &3.6               &                                                            &52.0\\
\object{IGR\,J17544$-$2619}     &O9 Ib$^2$         &2-4$^2$        &     0&06   &4.2--24&8+0&2.5 (2--4.3)   &                                                            &127.0\\
\object{XTE\,J1739$-$302}         &O8.5 Iab(f)$^3$&1.8-2.9$^3$ &     0&08   &3.0--28&12+1 &4.2 (2--8)   &50                                                       &126.4\\
\object{SAX\,J1818.6$-$1703}   &O9-B1 I$^4$     &                      &     0&18   &5.2--45&11+0    &2.9 (2--6)    &                                                          &76.9\\
\object{IGR\,J16479$-$4514}     &OB I$^5$           &                       &     0&2   &2.5--19&27+11 &3.6 (2--14)  &(16--35) + 84                                    &67.0\\
\object{AX\,J1841.0$-$0536}      &B0 I$^6$           &                        &$<0$&1&3.7--15&4+0     &5.8 (2--13.1)&                                                          &51.9\\
\object{AX\,J1820.5$-$1434}      &B$^4$               &                       &$<0$&1&3.4--5.3&4+0     &3.9 (2--9.6) &                                                            &59.4\\
\noalign{\vspace{2mm} \bf Intermediate systems\vspace{2mm}}
\object{IGR\,J16465$-$4507}     &B0.5 I$^7$        &                       &     0&1  &2.5--6.9&0+3    &                       &19, 25, 45                                          &66.7\\
\object{AX\,J1845.0$-$0433}       &O9 Ia$^8$        &3.6$^8$        &    0&2  &4.0--6.2 &6+0    &4.0 (2--14.3)&                                                            &55.2\\
\object{IGR\,J16195$-$4945}       &OB I$^9$         &7$^9$            &    0&2  &2.8--4.8 &6+0    &2.2 (2--3.3)  &                                                            &71.8\\
\object{IGR\,J16207$-$5129}       &B0 I$^4$          &4$^4$            &    0&4  &2.8--9.2 &9+2    &4.3 (2--11)   &18, 25                                                 &73.7\\
\object{XTE\,J1743$-$363}           &                         &                       &     0&5  &4.1--9.2 &16+3\,~ &2.5 (2--6.6)&21, 45, 61                                          &122.9\\
\noalign{\smallskip\hrule\smallskip}
\end{tabular}
Ref.: $^1$\cite{Leyder2007}, $^2$\cite{Pellizza2006}, $^3$\cite{Negueruela2006a}, $^4$\cite{Negueruela2007b}, $^5$\cite{Chaty2007}, $^6$\cite{Nespoli2007}, $^7$\cite{Negueruela2005}, $^8$\cite{Coe1996}, $^9$\cite{Sidoli2005}.
\end{table*}

\subsection{INTEGRAL data and analysis}

The sources of the sample are located along the galactic plane that has been heavily observed by INTEGRAL. All public data available in March 2007 are considered in this study. The INTEGRAL imager IBIS has a large field of view (FOV) of 29$\degr$ square with a good spatial resolution of 12$\arcmin$. Several thousands pointings including at least one source of the sample in its FOV and with an exposure longer than 600~s were selected, spanning times between Jan.~11, 2003 (revolution 30, MJD~52650) and Dec.~2, 2005 (revolution 383, MJD~53709). The total elapsed exposure time on each source goes from 52 to 127 days. The observations are not equally distributed in time due to scheduling reason.

The ISGRI data are reduced using the INTEGRAL Offline Scientific Analysis (OSA)
version 6.0 software publicly released by the INTEGRAL Science Data Centre \citep{courvoisier03AA}. Individual sky images for each pointing have been produced in the energy band 22--50~keV. The detection of the sources of the sample is forced in each individual images and the source count rate extracted. 

Sky maps gathering all pointings within the same revolution are also built. To build mosaic images with longer exposure time, each source is treated separately. Sky mosaics are built combining consecutive pointings in which the source is not detected at least at a 5.1$\sigma$ level. In several cases, when the source is not detected in several revolutions that are close in time, they are also mixed together to create one deep-exposure mosaic. Finally, one last mosaic is created combining all pointings where the source is not detected in order to constrain the lowest hard X-ray emission level. The light curves are built from the sky images considering only the pointings where the source is located less than 14$\degr$ away from the FOV centre.

The source count rate is extracted with the tool {\tt mosaic\_spec} (version 1.4) that is part of the OSA package. Detection of the source in mosaics are considered at a 6$\sigma$ level or higher. Therefore, each source light curve is built mixing short-- and long--term activity. The 22--50~keV count rates can be converted into physical fluxes with the relation $1\,\mathrm{Crab}=117\,\rm{ct/s} \simeq  9\times 10^{-9}\,\rm{erg}\,\rm{cm}^{-2}\,\rm{s}^{-1}$.

Because of the INTEGRAL dithering operations, the source could move out of the FOV and returns a few pointings later and still be detected. Source flares have therefore been detected by requiring a minimum of 25 ks of inactivity between them. Examples of flare lightcurves are presented in \cite{Sguera2006} and \cite{zurita2007}.

Flare duration of the order of a single INTEGRAL pointing $(2~\rm{ks})$ have been observed in all sources. (excepting \object{IGR\,J16465$-$4507}, but see below). Their typical duration is 3 ks. 

Fewer longer $(> 15~\rm{ks})$ flares have also been detected. As many flares are detected close to the sensitivity limit of the instrument, the flare duration is often difficult to evaluate and long flare could often also be considered as a series of shorter flares. We inspected each long flare carefully and a number of cases were found:

\begin{itemize}
\item In several sources (IGR J17544$-$2619, XTE J1739$-$ 302, SAX J1818.6$-$1703) some long flares were found to be composed of very strong short flares surrounded by activity at much lower level. Such flares were finally accounted in table \ref{tab2} as one short flare. Interestingly the detailed analysis of a flare of IGR J08408$-$4503 \citep{Leyder2007} also revealed that it was surrounded by emission at a lower level during several hours. This could be a common behavior, only detected by INTEGRAL for the brightest flares. 

\item In  XTE J1739$-$302 and SAX J1818.6$-$1703 a few long flares were clearly the result of two short flares separated by less than 25 ks. Such flares were accounted as two short flares in table \ref{tab2}.

\item In IGR J16207$-$5129, XTE 1743$-$363 and IGR J16465$-$4507, long flares were found to be sequences of detection in individual pointings, at the limit of detectability, separated by pointings where the source was in the FOV but too weak to be detected. When detected, the source count rate was similar to the sensitivity limit at the source off-axis angle. We conclude that those flares are particularly long periods of activity at the limit of the instrumental sensitivity. Such cases were kept in the ``long flare'' category in table \ref{tab2}. These sources are likely classical sgHMXB, with average fluxes just below the IBIS sensitivity. Such long activity periods cannot be considered as real flares.

\item IGR J16479-4514 is very active with one flare detected per day on average. 11 long flares were detected. 8 of them feature few very significant detections at pointing (2 ks) level separated by periods of $10-20~\rm{ks}$ during which the source was not detected. These flares could in fact be split in shorter flares with duration $\lesssim 8~\rm{ks}$. Three long flares are detected close to the detection limit and could also be represented as sequences of shorter flares. The large number of long flares detected in this source is therefore probably spurious and related to the high flare frequency. As a detailed analysis of the numerous flares of this source is beyond the scope of this paper we kept these long flares separated in table \ref{tab2}. 
\end{itemize}

We only have one really long flare remaining. A 50 ks flare has been detected in XTE J1739$-$302 as a continuous period of activity. It is very structured and made of at least 3 different peaks. It could be interpreted as a close sequence of shorter flares. 

\subsection{Variability and Source Classification}

Table \ref{tab2} lists the sources together with their quiescent count rate $(F_{q})$, typical flare count rate $(F_{fl})$, number of short ($<15$ ks) and long ($>15$ ks) flares $(N_{fl})$, range of flare durations $(t_{fl})$ for short and long flares and total source observing elapsed time $(T_{obs})$. The sources are ordered from high to low variability factor $(F_{fl}/F_{q})$.

The source observing time $T_{obs}$ is the sum of the elapsed time of all pontings with the source within $14\degr$ of the FOV center. As the instrument effective area decreases between the borders of the fully and partially coded fields of view, the probability to detect a flare effectively decreases when the source gets outside of the fully coded field of view. The effective observing time for flare detection can be estimated as $0.6~T_{obs}$. The effective time period between two flares is typically $T=7~\rm{days}$ on average but varies between 1 day in IGR J16479$-$4514 and two weeks in IGR J08408$-$4503 (this is uncertain as only two flares were detected).

The flare count rates $(F_{fl})$ are averaged over the duration of each flare. Peak count rates could be significantly larger than listed in table \ref{tab2}, especially for flares shorter than the pointing duration \citep[see e.g.][]{Leyder2007}.

The sources have been separated in two categories. The SFXT includes systems featuring hard X-ray variability by a factor $\gtrsim100$. ``Intermediate'' systems are candidate SFXT with smaller variability factors that could be compared with those of classical systems. The dividing line between the two source categories is not very well defined. From the variability point of view, sources closer to the bottom of the table are more similar to classical sgHMXB.

\section{Discussion}

\subsection{Hard X-ray luminosities}

The distance to the SFXT systems has been evaluated in a few cases \citep{Leyder2007, Negueruela2006a, Pellizza2006}. They range between 2 to 7 kpc with large uncertainties. We will assume, for the rest of the discussion a distance of 3 kpc. The average count rate observed during flares lies between 3 and 60 ct/s which translates to hard X-ray luminosities of $(0.2-4)\times 10^{36}~\rm{erg/s}$. Such luminosities are not exceptional for sgHMXB but very significantly larger than the typical X-ray luminosity of single massive stars of $10^{30-33}~\rm{erg/s}$ at soft X-rays \citep{Cassinelli1981}. The observed non thermal hard X-ray emission is therefore characteristic of accretion on a putative compact object and is, for now, the main reason to assume that SFXT are binary systems.

As the sources are flaring at most once per day, their average hard X-ray luminosity is very low, reaching $(0.2-4)\times 10^{34}~\rm{erg/s} $. It is therefore very unlikely that those systems have average orbital radius lower than $10^{13}~\rm{cm}$ i.e. $\sim 10~R_*$. One expects orbital periods larger than 15 days and underflow Roche lobe systems (note that no orbital period has yet been derived in any of these systems). 

The average hard X-ray luminosity of the SFXT systems in quiescence is $< 6.6 \times 10^{33}~\rm{erg/s}$ (which corresponds to 0.2 ct/s). This is an upper limit as the mosaics used to measure those quiescent fluxes most probably contain faint flares, not detected during single pointings. The INTEGRAL data alone do not exclude that there is no quiescent hard X-ray emission in these systems.

\subsection{Wind clumps\label{clump}}

The interaction of a compact object with a dense clump formed in the wind of a massive companion leads to increased accretion rate and hard X-ray emission \citep{Leyder2007}. 

The free-fall time from the accretion radius $R_a = 2\times 10^{10}~ \rm{cm}$ towards the compact object is of the order of $(2-3)\times10^2~\rm{s}$. As the intrinsic angular momentum of the accreted gas is small \citep{Illarionov2001} the infall is mostly radial (down to the Compton radius) and proceeds at the Bondi-Hoyle accretion rate. 

The accretion could slow down if the wind clumps have internal turbulence or harbor significant intrinsic angular momentum \citep{Theuns1996,Krumholz2005}. This is however unlikely in a highly supersonic wind, 
and supported by the very sharp X-ray flare cutoff observed on time scale of the order of few 100 s in some of these systems \citep{ZuritaWalter2007, gonzalez04aa}.

With a duration of $t_{fl}=2-10$ ks, the observed short hard X-ray flares are significantly longer than the free-fall time. The flare duration is therefore very probably linked with the thickness of the clumps which,  for a clump radial velocity $V_{cl}=10^8 ~\rm{cm/s}$, is $h_{cl} = V_{cl} \times t_{fl} \sim (2-10) \times 10^{11}~\rm{cm}$.

The average hard X-ray luminosity resulting from an interaction between the compact object and the clump can be evaluated as $L_X  = \epsilon~M_{acc}c^2/t_{fl}$ (where $\epsilon\sim0.1$) and the mass of a clump can then be estimated as
$$ M_{cl} = ~ (R_{cl}/R_{a})^2 ~M_{acc}= (R_{cl}/R_{a})^2~L_X~t_{fl}/(\epsilon~ c^2) $$
where $R_{cl}$ is the radius of the clump perpendicular to the radial distance.
In the case of a spherical clump,
$$M_{cl} = 
\left(\frac{L_X}{10^{36}~\rm{erg/s}}\right) \left(\frac{t_{fl}}{3~\rm{ks}}\right)^3 
~7.5\times 10^{21} ~\rm{g}.$$

If $\dot{N}$ is the rate of clumps emitted by the star, the observed hard X-ray flare rate is given by $T^{-1} = \dot{N}(R_{cl}^2/4R_{orb}^2).$
The rate of mass-loss in the form of wind clumps can then be estimated as
\begin{align}
\dot{M}_{cl} & = M_{cl}\dot{N}= \frac{4R_{orb}^2~L_X~t_{fl}}{R_{a}^2~T~\epsilon~ c^2}\notag\\
&={\left(\frac{10\rm{d}}{T}\frac{L_X}{10^{36}\rm{erg/s}}\frac{t_{fl}}{3\rm{ks}}\right)\left(\frac{R_{orb}}{10^{13}\rm{cm}}\right)^{2}} ~3\times 10^{-6}~\rm{M_{\odot}/y}.\notag
\end{align}

For a $\beta=1$ velocity law and spherical clumps, the number of clumps located between $1.05R_*$ and $R_{orb}$  can be evaluated as  
\begin{align}
N&=\dot{N}~
(t(R_{orb})-t(1.05R_*))\notag\\
&=\left(\frac{10~\rm{d}}{T}\right)\left(\frac{3~\rm{ks}}{t_{fl}}\right)^2\left( \frac{R_{orb}}{10^{13}~\rm{cm}}\right)^3~3.8\times 10^3,\notag
\end{align}
where t(r) is the wind flight time \citep{Hamann2001}. 

Assuming spherical clumps, the clump density at the orbital radius is $\rho_{cl}=\left(\frac{L_X}{10^{36}~\rm{erg/s}}\right) ~7\times 10^{-14} ~\rm{g~cm}^{-3}$ and the corresponding homogeneous wind density is $\rho_h=\dot{M}_{cl}/(4\pi~R_{orb}^2~V_{cl})=
\left(\frac{10~\rm{d}}{T}\frac{L_X}{10^{36}~\rm{erg/s}}\frac{t_{fl}}{3~\rm{ks}}\right)
~1.5\times 10^{-15}~\rm{g~cm}^{-3}$. The clump volume filling factor at the orbital radius is $
f_V = \frac{\rho_h}{\rho_{cl}} = 
\left(\frac{10~\rm{d}}{T}\frac{t_{fl}}{3~\rm{ks}}\right)
~0.02$ and the corresponding porosity length \citep{owocki2006,OskinovaHamannFeldmeier2007} is
$h=\frac{R_{cl}}{f_V}=
\left(\frac{T}{10~\rm{d}}\right)
~15\times 10^{12} ~\rm{cm}$.

If the density of a clump decreases with radius as $r^{-2\beta}$ and its mass remains constant, the averaged homogeneous wind density within $R_{obs}$ is  $\overline{\rho_{h}}=N M_{cl}/(\frac{4}{3}\pi 
R_{orb}^3
) = 
\left(\frac{10~\rm{d}}{T}\frac{L_X}{10^{36}~\rm{erg/s}}\frac{t_{fl}}{3~\rm{ks}}\right)
~7\times 10^{-15} ~\rm{g~cm}^{-3}$ and the average clump volume filling factor and porosity length could be estimated as 0.1 and $3\times10^{12} ~\rm{cm}$, respectively.

The variety of $t_{fl}$, $T$ and $F_{fl}$ that are observed probably reflects a range of  clump parameters and orbital radii. Several of the average clump parameters estimated above, in particular the clump density, filling factor and  porosity length do not depend on the orbital radius, which is unknown, and only slowly depend on the observed quantities.
 
These average parameters match the macro-clumping scenario proposed by \cite{OskinovaHamannFeldmeier2007}
to reconcile clumping and mass-loss rates. Their model depends on a free parameter $L_0=L(r) (r^2V(r)/V(\infty))^{-1/3}$, where $L(r)$ is the clump separation in unit of $R_*$. Our average clumping parameters correspond to $L_0=0.35$. The number of clumps derived above is also comparable to evaluations by \cite{Lepine1999} and \cite{OskinovaFeldmeierHamann2006, OskinovaHamannFeldmeier2007}. The volume filling factor and the clump mass-loss rate are also similar to those derived by \cite{Bouret2005} from the study of ultraviolet and optical line profiles in two super-giant stars.

The column density through a clump can also be estimated as $N_H = \frac{M_{cl}}{R_{cl}^2m_p}=
\left(\frac{L_X}{10^{36}\rm{erg/s}}\frac{t_{fl}}{3\rm{ks}}\right)
~5\times 10^{22}\rm{cm}^{-2}$. The clumps remain optically thin in the X-rays.

The activity surrounding some of the most significant flares may be related to tidal effects on the clumps themself and induced turbulence. The long flare that has been observed in XTE J1739$-$302 is structured with many sub-peaks and may correspond to the clustering of smaller wind clumps perhaps related to the dissolution of pancake or shell shaped wind clumps. 

\subsection{Missed flares}

The mass spectrum of the wind clumps is completely unknown. Only the brightest hard X-ray flares in SFXT and ``intermediate'' objects could have been detected by INTEGRAL. Bright and short as well as long and faint flares could have remained undetected.

By definition, the contribution of missed flare to the average source count rate could not be differentiated from the quiescent emission. 
The average contribution of undetected flare, which is not larger than the quiescent count rate listed in table \ref{tab2}, can be compared to the contribution of the detected flares.

In two SFXT (IGR J16479$-$4514, AX J1820.5$-$1434) the average count rate is more than 3 times larger than the quiescent count rate. In these cases, the global contribution of undetected flares is at most a third of the contribution of the detected flares. In three additional sources (IGR J16465$-$4507, XTE J1739$-$302, AXJ 1841.0$-$0536) the average count rate is slightly larger than the quiescent count rate and the global contribution of undetected flares is at most similar to the contribution of the detected flares. In these 5 sources, INTEGRAL thus detected a significant fraction of the accretion luminosity.

In the two remaining SFXT (IGR J17544$-$2619, SAX J1818.6$-$1703) and in the other ``intermediate'' sources, the average count rate is compatible with the quiescent count rate. One cannot therefore exclude that the contribution of undetected flares to the average hard X-ray luminosity could be larger than that of detected flares. 

A large fraction of the hard X-ray emission detected in SFXT is therefore generated by the accretion of the clumps described in Sect. \ref{clump}. 

A population of small clumps (with $t_{fl}<300~\rm{s}$) could have remained almost undetected if their interaction luminosity is not larger than $10^{36}~\rm{erg/s}$. Their mass would be of the order of $10^{19}~\rm{g}$. Short flares would need to be very frequent $(T\lesssim 1~\rm{day})$ for those clumps to contribute significantly to the mass-loss rate and should be easily detectable in X-ray observations of a sample of sources.

Undetected long ($t_{fl}>50~\rm{ks}$) and faint $(L_{X}\lesssim 10^{35}~\rm{erg/s})$ flares would be the signature of extremely large $(R_{cl}\gtrsim R_*)$ and massive $(10^{24}~\rm{g})$ clumps with a density of the order of the equivalent homogeneous wind density. Such structures would be interpreted as wind in-homogeneities with low mass-loss rate rather than by clumping.

\subsection{Inter-clump medium}

The variation of the observed X-ray flux between flares and quiescence provides in principle a direct measure of the density constrast between the wind clumps and the inter-clump medium. As part of the quiescent count rate could be related to undetected flares or to other emission, these density contrasts are lower limits.

As the hard X-ray flux variability observed in SFXT (table \ref{tab2}) reaches factors $> 100$ when a quiescent count rate could be firmly detected with INTEGRAL, the density contrast is at least of a factor 100. In the ``intermediate'' sources INTEGRAL observed variability factors larger than 20. 
Additional constraints are available for the sources observed at higher sensitivity in the X-rays:
\begin{itemize}
\item In IGR J08408$-$4503, \cite{Leyder2007} have analysed SWIFT XRT observations which provided a very faint quiescent luminosity of $2\times 10^{32}~\rm{erg/s}$ in the form of a soft, probably thermal, spectrum. In this source the wind density contrast was larger than $10^4$.

\item The flaring activity of IGR J17544$-$2619 was observed by XMM/Newton \citep{gonzalez04aa}. These flares were short and very structured on time scale of 100 s. The variation of the count rates observed between quiescence and average or peak flare are of a factor 20 to 40. Similar and stronger variations have been observed by INTEGRAL \citep{walter2006}. 

\item IGR J16479$-$4514 was observed by XMM/Newton and it was found 20 times fainter than the average flux observed by INTEGRAL  \citep{walter2006}. The overall variability observed in this source is of the order of 100. 

\item In AX J1845.0$-$0433, the flare and quiescent fluxes were observed simultaneously by XMM-Newton. The quiescent spectrum is clearly dominated by accretion \citep{ZuritaWalter2007}.  A density contrast of 15-25 was detected.

\item IGR J16465$-$4507 was observed by XMM-Newton \citep{walter2006} and IGR J16195$-$4945, IGR J16207$-$5129 \citep{Tomsick2006}, AX J1841.0$-$0536 \citep{Halpern2004}, SAX J1818$-$1703 \citep{intZand2006} and XTE 1739-302 \citep{Smith2006} were observed by Chandra. In all cases, the X-ray flux is similar to the average quiescent fluxes observed by INTEGRAL (table \ref{tab2}). \end{itemize}

Density contrasts of $>10^{2-4}$ and 15--50 have been observed in SFXT and ``Intermediate'' sources, respectively. The density contrast is larger in SFXT than in ``Intermediate'' and, of course, classical systems. Density contrasts are probably stronger when clumping is very effective. 

Numerical simulations of the line driven instability \citep{Runacres2005} predict density contrasts as large as $10^{3-5}$ in the wind up to large radii. At a distance of $10~R_*$, the simulated density can vary between $10^{-18}$ and $10^{-13}~\rm{g~cm^{-3}}$ and the separation between the density peaks are of the order of  $R_*$. These characteristics are comparable to the values we have derived from the high-energy observations.

\subsection{Where do classical sgHMXB fit in this picture ?}

Classical sgHMXB are characterized by small orbital radii $R_{orb}=(1.5-2.5)~R_*$, and by flux variability of a factor $\lesssim20$. Such variabilities were modelled in term of wind inhomogeneities largely triggered by the hydrodynamic and photo-ionisation effects of the accreting object on the companion and inner stellar wind \citep{blondin91, blondin94}. At small orbital radii, the companion is close to fill its Roche lobe, which triggers tidal streams. In addition the X-ray source ionizes the wind acceleration zone, prevents wind acceleration and generates slower velocities, denser winds, larger accretion radius and finally larger X-ray luminosities. Whether or not the stellar wind is intrinsically clumpy at low radius, the effect of the compact object on the wind is expected to be important.

The main difference between SFXT and classical sgHMXB could therefore be their orbital radius. At very low orbital radius $(<1.5~R_*)$ tidal accretion will take place through an accretion disk and the system will soon evolve to a common envelope stage. At low orbital radius $(\sim 2~R_*)$ the wind will be perturbed in any case and efficient wind accretion will lead to copious and persistent X-ray emission $(10^{36-37}~\rm{erg/s})$. At larger orbital radius $(\sim 10~R_*)$ and if the wind is clumpy, the SFXT behavior is expected as described above. If the wind clumps do not form for any reason, the average accretion rate will remain too low and the sources will remain mostly undetected by the current hard X-ray surveying instrumentation.

\section{Conclusions}

INTEGRAL tripled the number of super-giant HMXB systems known in the Galaxy and revealed two new populations: the absorbed and the fast transient (SFXT) systems. The typical hard X-ray variability factor is $\lesssim 20$ in classical and absorbed systems and $\gtrsim 100$ in SFXT. We have also identified some ``intermediate'' systems with smaller variability factors that could be either SFXT or classical systems.

The SFXT behavior is best explained by the interaction between the accreting  compact object and a clumpy stellar wind \citep{IntZand2005,Leyder2007}. Using the hard X-ray variability observed by INTEGRAL in a sample of SFXT we have derived typical wind clump parameters. The compact object orbital radius are probably relatively large ($10~R_*$) and the clumps which generate most of the hard X-ray emission have a size of a few tenth of $R_*$. The clump mass is of the order or $10^{22-23}~\rm{g}$ (for a column density of $10^{22-23}~\rm{cm}^{-2}$) and the corresponding mass-loss rate is $10^{-(5-6)}~\rm{M_{\odot}/y}$. At the orbital radius, the clump separation is of the order of $R_*$ and their volume filling factor is $0.02$. Depending how the clump density varies with radius, the average volume filling factor could be as large as 0.1.

These parameters are in good agreement with the macro clumping scenario proposed by \cite{OskinovaHamannFeldmeier2007}.

The observed ratio between the flare and quiescent count rates indicate density ratios between the clumps and the inter-clump medium which vary between 15 to 50 in ``Intermediate'' systems and $10^{2-4}$ in SFXT. Such ratios and the observed clump densities are in reasonable agreement with the predictions of line driven instabilities at large radii \citep{Runacres2005}.

The main difference between classical sgHMXB and SFXT could be the orbital radius of the compact object. At small orbital radius ($R_{orb}\sim2~R_*$) the systems are persistent and luminous. At larger radius and if wind clumping takes place the fast transient SFXT behavior is observed.

\begin{acknowledgements}
Based on observations with INTEGRAL, an ESA project with instruments and science data centre funded by ESA member states (especially the PI countries: Denmark, France, Germany, Italy, Switzerland, Spain), Czech Republic and Poland, and with the participation of Russia and the USA.\end{acknowledgements}

\bibliographystyle{aa}
\bibliography{8353}

\begin{thebibliography}{63}
\expandafter\ifx\csname natexlab\endcsname\relax\def\natexlab#1{#1}\fi

\bibitem[{{Abbott} {et~al.}(1981){Abbott}, {Bieging}, \&
  {Churchwell}}]{Abbott1981}
{Abbott}, D.~C., {Bieging}, J.~H., \& {Churchwell}, E. 1981, \apj, 250, 645

\bibitem[{{Blondin}(1994)}]{blondin94}
{Blondin}, J.~M. 1994, \apj, 435, 756

\bibitem[{{Blondin} {et~al.}(1991){Blondin}, {Stevens}, \&
  {Kallman}}]{blondin91}
{Blondin}, J.~M., {Stevens}, I.~R., \& {Kallman}, T.~R. 1991, \apj, 371, 684

\bibitem[{{Bodaghee} {et~al.}(2006){Bodaghee}, {Walter}, {Zurita Heras},
  {Bird}, {Courvoisier}, {Malizia}, {Terrier}, \& {Ubertini}}]{bodaghee05}
{Bodaghee}, A., {Walter}, R., {Zurita Heras}, J.~A., {et~al.} 2006, \aap, 447,
  1027

\bibitem[{{Bouret} {et~al.}(2005){Bouret}, {Lanz}, \& {Hillier}}]{Bouret2005}
{Bouret}, J.-C., {Lanz}, T., \& {Hillier}, D.~J. 2005, \aap, 438, 301

\bibitem[{{Cassinelli} \& {Olson}(1979)}]{CassinelliOlson1979}
{Cassinelli}, J.~P. \& {Olson}, G.~L. 1979, \apj, 229, 304

\bibitem[{{Cassinelli} {et~al.}(1981){Cassinelli}, {Waldron}, {Sanders},
  {Harnden}, {Rosner}, \& {Vaiana}}]{Cassinelli1981}
{Cassinelli}, J.~P., {Waldron}, W.~L., {Sanders}, W.~T., {et~al.} 1981, \apj,
  250, 677

\bibitem[{{Castor} {et~al.}(1975){Castor}, {Abbott}, \& {Klein}}]{cak1975}
{Castor}, J.~I., {Abbott}, D.~C., \& {Klein}, R.~I. 1975, \apj, 195, 157

\bibitem[{{Chaty} \& {al.}(2007)}]{Chaty2007}
{Chaty}, S. \& {al.} 2007, \aap, in preparation

\bibitem[{{Coe} {et~al.}(1996){Coe}, {Fabregat}, {Negueruela}, {Roche}, \&
  {Steele}}]{Coe1996}
{Coe}, M.~J., {Fabregat}, J., {Negueruela}, I., {Roche}, P., \& {Steele}, I.~A.
  1996, \mnras, 281, 333

\bibitem[{{Courvoisier} {et~al.}(2003){Courvoisier}, {Walter}, {Beckmann},
  {Dean}, {Dubath}, {Hudec}, {Kretschmar}, {Mereghetti}, {Montmerle},
  {Mowlavi}, {Paltani}, {Preite Martinez}, {Produit}, {Staubert}, {Strong},
  {Swings}, {Westergaard}, {White}, {Winkler}, \&
  {Zdziarski}}]{courvoisier03AA}
{Courvoisier}, T.~J.-L., {Walter}, R., {Beckmann}, V., {et~al.} 2003, \aap,
  411, L53

\bibitem[{{Davies} {et~al.}(2007){Davies}, {Vink}, \& {Oudmaijer}}]{Davies2007}
{Davies}, B., {Vink}, J.~S., \& {Oudmaijer}, R.~D. 2007, \aap, 469, 1045

\bibitem[{{Feldmeier}(1995)}]{feldmeier1995}
{Feldmeier}, A. 1995, \aap, 299, 523

\bibitem[{{Fullerton} {et~al.}(2006){Fullerton}, {Massa}, \&
  {Prinja}}]{Fullerton2006}
{Fullerton}, A.~W., {Massa}, D.~L., \& {Prinja}, R.~K. 2006, \apj, 637, 1025

\bibitem[{{Gonz{\' a}lez-Riestra} {et~al.}(2004){Gonz{\' a}lez-Riestra},
  {Oosterbroek}, {Kuulkers}, {Orr}, \& {Parmar}}]{gonzalez04aa}
{Gonz{\' a}lez-Riestra}, R., {Oosterbroek}, T., {Kuulkers}, E., {Orr}, A., \&
  {Parmar}, A.~N. 2004, \aap, 420, 589

\bibitem[{{Halpern} {et~al.}(2004){Halpern}, {Gotthelf}, {Helfand}, {Gezari},
  \& {Wegner}}]{Halpern2004}
{Halpern}, J.~P., {Gotthelf}, E.~V., {Helfand}, D.~J., {Gezari}, S., \&
  {Wegner}, G.~A. 2004, The Astronomer's Telegram, 289, 1

\bibitem[{{Hamann} {et~al.}(2001){Hamann}, {Brown}, {Feldmeier}, \&
  {Oskinova}}]{Hamann2001}
{Hamann}, W.-R., {Brown}, J.~C., {Feldmeier}, A., \& {Oskinova}, L.~M. 2001,
  \aap, 378, 946

\bibitem[{{Hill} {et~al.}(2005){Hill}, {Walter}, {Knigge}, {Bazzano},
  {B{\'e}langer}, {Bird}, {Dean}, {Galache}, {Malizia}, {Renaud}, {Stephen}, \&
  {Ubertini}}]{hill05aa}
{Hill}, A.~B., {Walter}, R., {Knigge}, C., {et~al.} 2005, \aap, 439, 255

\bibitem[{{Hirschi}(2007)}]{Hirschi2007}
{Hirschi}, R. 2007, in Proceedings of the workshop ``Clumping in Hot-Star
  Winds", Potsdam, Germany, 18.-22. June 2007, Universit\"atsverlag Potsdam

\bibitem[{Illarionov \& Beloborodov(2001)}]{Illarionov2001}
Illarionov, A.~F. \& Beloborodov, A.~M. 2001, Mon. Not. Roy. Astron. Soc., 323,
  159

\bibitem[{{in't Zand} {et~al.}(2006){in't Zand}, {Jonker}, {Mendez}, \&
  {Markwardt}}]{intZand2006}
{in't Zand}, J., {Jonker}, P., {Mendez}, M., \& {Markwardt}, C. 2006, The
  Astronomer's Telegram, 915, 1

\bibitem[{{in't Zand}(2005)}]{IntZand2005}
{in't Zand}, J.~J.~M. 2005, \aap, 441, L1

\bibitem[{{in't Zand} {et~al.}(2007){in't Zand}, {Kuiper}, {den Hartog},
  {Hermsen}, \& {Corbet}}]{Zand2007}
{in't Zand}, J.~J.~M., {Kuiper}, L., {den Hartog}, P.~R., {Hermsen}, W., \&
  {Corbet}, R.~H.~D. 2007, \aap, 469, 1063

\bibitem[{{Krumholz} {et~al.}(2005){Krumholz}, {McKee}, \&
  {Klein}}]{Krumholz2005}
{Krumholz}, M.~R., {McKee}, C.~F., \& {Klein}, R.~I. 2005, \apj, 618, 757

\bibitem[{{Lebrun} {et~al.}(2003){Lebrun}, {Leray}, {Lavocat}, {Cr{\' e}tolle},
  {Arqu{\` e}s}, {Blondel}, {Bonnin}, {Bou{\` e}re}, {Cara}, {Chaleil}, {Daly},
  {Desages}, {Dzitko}, {Horeau}, {Laurent}, {Limousin}, {Mathy}, {Mauguen},
  {Meignier}, {Molini{\' e}}, {Poindron}, {Rouger}, {Sauvageon}, \&
  {Tourrette}}]{lebrun03AA}
{Lebrun}, F., {Leray}, J.~P., {Lavocat}, P., {et~al.} 2003, \aap, 411, L141

\bibitem[{{L{\'e}pine} \& {Moffat}(1999)}]{Lepine1999}
{L{\'e}pine}, S. \& {Moffat}, A.~F.~J. 1999, \apj, 514, 909

\bibitem[{{Leyder} {et~al.}(2007){Leyder}, {Walter}, {Lazos}, {Masetti}, \&
  {Produit}}]{Leyder2007}
{Leyder}, J.-C., {Walter}, R., {Lazos}, M., {Masetti}, N., \& {Produit}, N.
  2007, \aap, 465, L35

\bibitem[{{Lupie} \& {Nordsieck}(1987)}]{Lupie1987}
{Lupie}, O.~L. \& {Nordsieck}, K.~H. 1987, \aj, 93, 214

\bibitem[{{Markova} {et~al.}(2005){Markova}, {Puls}, {Scuderi}, \&
  {Markov}}]{Markova2005}
{Markova}, N., {Puls}, J., {Scuderi}, S., \& {Markov}, H. 2005, \aap, 440, 1133

\bibitem[{{Negueruela} \& {Schurch}(2007)}]{Negueruela2007b}
{Negueruela}, I. \& {Schurch}, M.~P.~E. 2007, \aap, 461, 631

\bibitem[{{Negueruela} {et~al.}(2005){Negueruela}, {Smith}, \&
  {Chaty}}]{Negueruela2005}
{Negueruela}, I., {Smith}, D.~M., \& {Chaty}, S. 2005, The Astronomer's
  Telegram, 429, 1

\bibitem[{{Negueruela} {et~al.}(2006{\natexlab{a}}){Negueruela}, {Smith},
  {Harrison}, \& {Torrej{\'o}n}}]{Negueruela2006a}
{Negueruela}, I., {Smith}, D.~M., {Harrison}, T.~E., \& {Torrej{\'o}n}, J.~M.
  2006{\natexlab{a}}, \apj, 638, 982

\bibitem[{{Negueruela} {et~al.}(2006{\natexlab{b}}){Negueruela}, {Smith},
  {Reig}, {Chaty}, \& {Torrej{\'o}n}}]{Negueruela2006}
{Negueruela}, I., {Smith}, D.~M., {Reig}, P., {Chaty}, S., \& {Torrej{\'o}n},
  J.~M. 2006{\natexlab{b}}, in ESA Special Publication, Vol. 604, The X-ray
  Universe 2005, ed. A.~{Wilson}, 165--170

\bibitem[{{Negueruela} {et~al.}(2007){Negueruela}, {Smith}, {Torrejon}, \&
  {Reig}}]{Negueruela2007}
{Negueruela}, I., {Smith}, D.~M., {Torrejon}, J.~M., \& {Reig}, P. 2007, ArXiv
  e-prints astro-ph/07043224

\bibitem[{{Nespoli} {et~al.}(2007){Nespoli}, {Fabregat}, \&
  {Mennickent}}]{Nespoli2007}
{Nespoli}, E., {Fabregat}, J., \& {Mennickent}, R. 2007, The Astronomer's
  Telegram, 983, 1

\bibitem[{{Oskinova} {et~al.}(2006){Oskinova}, {Feldmeier}, \&
  {Hamann}}]{OskinovaFeldmeierHamann2006}
{Oskinova}, L.~M., {Feldmeier}, A., \& {Hamann}, W.-R. 2006, \mnras, 372, 313

\bibitem[{{Oskinova} {et~al.}(2007){Oskinova}, {Hamann}, \&
  {Feldmeier}}]{OskinovaHamannFeldmeier2007}
{Oskinova}, L.~M., {Hamann}, W.-R., \& {Feldmeier}, A. 2007, ArXiv e-prints
  astro-ph/07042390

\bibitem[{{Owocki} {et~al.}(1988){Owocki}, {Castor}, \& {Rybicki}}]{ocr1988}
{Owocki}, S.~P., {Castor}, J.~I., \& {Rybicki}, G.~B. 1988, \apj, 335, 914

\bibitem[{{Owocki} \& {Cohen}(2006)}]{owocki2006}
{Owocki}, S.~P. \& {Cohen}, D.~H. 2006, \apj, 648, 565

\bibitem[{{Pellizza} {et~al.}(2006){Pellizza}, {Chaty}, \&
  {Negueruela}}]{Pellizza2006}
{Pellizza}, L.~J., {Chaty}, S., \& {Negueruela}, I. 2006, \aap, 455, 653

\bibitem[{{Pittard}(2007)}]{pittard2007}
{Pittard}, J.~M. 2007, \apjl, 660, L141

\bibitem[{{Pollock} {et~al.}(2005){Pollock}, {Corcoran}, {Stevens}, \&
  {Williams}}]{Pollock2005}
{Pollock}, A.~M.~T., {Corcoran}, M.~F., {Stevens}, I.~R., \& {Williams}, P.~M.
  2005, \apj, 629, 482

\bibitem[{{Prinja} \& {Howarth}(1986)}]{PrinjaHowarth1986}
{Prinja}, R.~K. \& {Howarth}, I.~D. 1986, \apjs, 61, 357

\bibitem[{{Rodriguez} {et~al.}(2006){Rodriguez}, {Bodaghee}, {Kaaret},
  {Tomsick}, {Kuulkers}, {Malaguti}, {Petrucci}, {Cabanac}, {Chernyakova},
  {Corbel}, {Deluit}, {Di Cocco}, {Ebisawa}, {Goldwurm}, {Henri}, {Lebrun},
  {Paizis}, {Walter}, \& {Foschini}}]{Rodriguez2006}
{Rodriguez}, J., {Bodaghee}, A., {Kaaret}, P., {et~al.} 2006, \mnras, 366, 274

\bibitem[{{Rodriguez} {et~al.}(2005){Rodriguez}, {Cabanac}, {Hannikainen},
  {Beckmann}, {Shaw}, \& {Schultz}}]{rodriguez05aa}
{Rodriguez}, J., {Cabanac}, C., {Hannikainen}, D.~C., {et~al.} 2005, \aap, 432,
  235

\bibitem[{{Runacres} \& {Owocki}(2005)}]{Runacres2005}
{Runacres}, M.~C. \& {Owocki}, S.~P. 2005, \aap, 429, 323

\bibitem[{{Sako} {et~al.}(2003){Sako}, {Kahn}, {Paerels}, {Liedahl},
  {Watanabe}, {Nagase}, \& {Takahashi}}]{Sako2003}
{Sako}, M., {Kahn}, S.~M., {Paerels}, F., {et~al.} 2003, ArXiv e-prints
  astro-ph/0309503, invited review at the High-resolution X-ray Spectroscopy
  Workshop with XMM-Newton and Chandra, MSSL, Oct 24-25, 2002

\bibitem[{{Schild} {et~al.}(2004){Schild}, {G{\"u}del}, {Mewe}, {Schmutz},
  {Raassen}, {Audard}, {Dumm}, {van der Hucht}, {Leutenegger}, \&
  {Skinner}}]{Schild2004}
{Schild}, H., {G{\"u}del}, M., {Mewe}, R., {et~al.} 2004, \aap, 422, 177

\bibitem[{{Sguera} {et~al.}(2006){Sguera}, {Bazzano}, {Bird}, {Dean},
  {Ubertini}, {Barlow}, {Bassani}, {Clark}, {Hill}, {Malizia}, {Molina}, \&
  {Stephen}}]{Sguera2006}
{Sguera}, V., {Bazzano}, A., {Bird}, A.~J., {et~al.} 2006, \apj, 646, 452

\bibitem[{{Sidoli} {et~al.}(2006){Sidoli}, {Paizis}, \&
  {Mereghetti}}]{Sidoli2006}
{Sidoli}, L., {Paizis}, A., \& {Mereghetti}, S. 2006, \aap, 450, L9

\bibitem[{{Sidoli} {et~al.}(2005){Sidoli}, {Vercellone}, {Mereghetti}, \&
  {Tavani}}]{Sidoli2005}
{Sidoli}, L., {Vercellone}, S., {Mereghetti}, S., \& {Tavani}, M. 2005, \aap,
  429, L47

\bibitem[{{Smith} {et~al.}(2006){Smith}, {Heindl}, {Markwardt}, {Swank},
  {Negueruela}, {Harrison}, \& {Huss}}]{Smith2006}
{Smith}, D.~M., {Heindl}, W.~A., {Markwardt}, C.~B., {et~al.} 2006, \apj, 638,
  974

\bibitem[{{Theuns} {et~al.}(1996){Theuns}, {Boffin}, \&
  {Jorissen}}]{Theuns1996}
{Theuns}, T., {Boffin}, H.~M.~J., \& {Jorissen}, A. 1996, \mnras, 280, 1264

\bibitem[{{Tomsick} {et~al.}(2006){Tomsick}, {Chaty}, {Rodriguez}, {Foschini},
  {Walter}, \& {Kaaret}}]{Tomsick2006}
{Tomsick}, J.~A., {Chaty}, S., {Rodriguez}, J., {et~al.} 2006, \apj, 647, 1309

\bibitem[{{Ubertini} {et~al.}(2003){Ubertini}, {Lebrun}, {Di Cocco}, {Bazzano},
  {Bird}, {Broenstad}, {Goldwurm}, {La Rosa}, {Labanti}, {Laurent}, {Mirabel},
  {Quadrini}, {Ramsey}, {Reglero}, {Sabau}, {Sacco}, {Staubert}, {Vigroux},
  {Weisskopf}, \& {Zdziarski}}]{ubertini03AA}
{Ubertini}, P., {Lebrun}, F., {Di Cocco}, G., {et~al.} 2003, \aap, 411, L131

\bibitem[{{van der Meer} {et~al.}(2005){van der Meer}, {Kaper}, {di Salvo},
  {M{\'e}ndez}, {van der Klis}, {Barr}, \& {Trams}}]{Vandermeer2005}
{van der Meer}, A., {Kaper}, L., {di Salvo}, T., {et~al.} 2005, \aap, 432, 999

\bibitem[{{Walter} {et~al.}(2004){Walter}, {Courvoisier}, {Foschini}, {Lebrun},
  {Lund}, {Parmar}, {Rodriguez}, {Tomsick}, \& {Ubertini}}]{walter04esasp}
{Walter}, R., {Courvoisier}, R., {Foschini}, L., {et~al.} 2004, in Proceedings
  of the 5th INTEGRAL workshop, 16-20 February 2004, Munich, Germany
  (ESA-SP-552), Vol. 552, 417--421

\bibitem[{{Walter} {et~al.}(2006){Walter}, {Zurita Heras}, {Bassani},
  {Bazzano}, {Bodaghee}, {Dean}, {Dubath}, {Parmar}, {Renaud}, \&
  {Ubertini}}]{walter2006}
{Walter}, R., {Zurita Heras}, J., {Bassani}, L., {et~al.} 2006, \aap, 453, 133

\bibitem[{{Winkler} {et~al.}(2003){Winkler}, {Courvoisier}, {Di Cocco},
  {Gehrels}, {Gim{\' e}nez}, {Grebenev}, {Hermsen}, {Mas-Hesse}, {Lebrun},
  {Lund}, {Palumbo}, {Paul}, {Roques}, {Schnopper}, {Sch{\" o}nfelder},
  {Sunyaev}, {Teegarden}, {Ubertini}, {Vedrenne}, \& {Dean}}]{winkler03AA}
{Winkler}, C., {Courvoisier}, T.~J.-L., {Di Cocco}, G., {et~al.} 2003, \aap,
  411, L1

\bibitem[{{Zurita Heras} \& {Chaty}(2007)}]{ZuritaChaty2007}
{Zurita Heras}, J.~A. \& {Chaty}, S. 2007, in preparation

\bibitem[{{Zurita Heras} {et~al.}(2007){Zurita Heras}, {Chaty}, \&
  {Rodriguez}}]{zurita2007}
{Zurita Heras}, J.~A., {Chaty}, S., \& {Rodriguez}, J. 2007, The Astronomer's
  Telegram, 1035, 1

\bibitem[{{Zurita Heras} {et~al.}(2006){Zurita Heras}, {de Cesare}, {Walter},
  {Bodaghee}, {B{\'e}langer}, {Courvoisier}, {Shaw}, \& {Stephen}}]{zurita05}
{Zurita Heras}, J.~A., {de Cesare}, G., {Walter}, R., {et~al.} 2006, \aap, 448,
  261

\bibitem[{{Zurita Heras} \& {Walter}(2007)}]{ZuritaWalter2007}
{Zurita Heras}, J.~A. \& {Walter}, R. 2007, \aap, submitted

\end{thebibliography}
\newpage

\begin{figure}[h]
\vfill
\end{figure}

\newpage

\end{document}